\documentclass[twocolumn,aps,amsmath,amssymb,showpacs,showkeys,floatfix,prl,superscriptaddress]{revtex4}
\usepackage{bbm}
\usepackage[dvips]{graphicx}
\usepackage{color}
\usepackage{amsmath,amssymb,latexsym}
\usepackage{upgreek}

\newcommand{\micron}{\ensuremath{\mu\mathrm{m}}}
\newcommand{\efficiency}{\eta}

\renewcommand{\P}{\mathbf{P}}
\newcommand{\Q}{\mathbf{Q}}
\newcommand{\q}{\mathbf{q}}

\newcommand{\Z}{Z}

\renewcommand{\ol}[1]{\overline{#1}}

\newcommand{\R}{\mathcal{R}}

\begin{document}

\title{The load-response of the flagellar beat \protect\\ \textit{(accepted for publication in Phys. Rev. Lett.)}}
\author{Gary S. Klindt}
\affiliation{Max Planck Institute for the Physics of Complex Systems, Dresden, Germany}
\author{Christian Ruloff}
\affiliation{Experimental Physics, University of the Saarland, Saarbr\"ucken, Germany}
% Experimental Physics, Saarland University, 66123 Saarbrücken, Germany
\author{Christian Wagner}
\affiliation{Experimental Physics, University of the Saarland, Saarbr\"ucken, Germany}
\affiliation{Physics and Materials Science Research Unit, University of Luxembourg, Luxembourg, Luxembourg}
% Physics and Materials Science Research Unit, University of Luxembourg, 162a avenue de la Faïencerie, L-1511 Luxembourg, Luxembourg
\author{Benjamin M. Friedrich}
\email{benjamin.m.friedrich@tu-dresden.de}
\affiliation{Max Planck Institute for the Physics of Complex Systems, Dresden, Germany}
\affiliation{Center for Advancing Electronics Dresden cfaed, TU Dresden, Dresden, Germany}

\date{\today}


 %    synchronization; coupled oscillators
% 47.63.Gd} %    Swimming microorganisms

\begin{abstract}
  Cilia and flagella exhibit regular bending waves that perform
  mechanical work on the surrounding fluid, to propel cellular
  swimmers and pump fluids inside organisms. Here, we quantify a
  force-velocity relationship of the beating flagellum, by exposing
  flagellated \emph{Chlamydomonas} cells to controlled microfluidic
  flows. A simple theory of
  flagellar limit-cycle oscillations, calibrated by measurements in
  the absence of flow, reproduces this relationship quantitatively. We
  derive a link between the energy efficiency of the
  flagellar beat and its ability to synchronize to oscillatory flows.
\end{abstract}

\maketitle

\paragraph{Introduction.}
Motile cilia and flagella are slender cell appendages of length
$10-100 \,\micron$ that exhibit regular bending waves with typical
frequencies of $10-100\,\mathrm{Hz}$~\cite{Gray:1928}, rendering the
beating flagellum a prime example of a micro-mechanical oscillator.
The flagellar beat propels eukaryotic microswimmers in a liquid, including
sperm cells, non-bacterial pathogens, and green alga. 
In the human body, self-organized metachronal waves of short flagella termed cilia
transport fluids, \textit{e.g.} mucus in the
airways~\cite{sanderson1981ciliary} and cerebrospinal fluid in the
brain~\cite{worthington1963ependymal}. The cytoskeletal core of cilia
and flagella, the axoneme, was highly conserved in evolution, from
protists to plants and animals~\cite{Carvalho-Santos2011}.

Flagellar bending waves arise by a dynamic instability in the
collective dynamics of $10^4-10^5$ molecular dynein motors, which are
regularly spaced along the flagellar axoneme~\cite{Nicastro2006}.
This motor activity empowers the eukaryotic flagellum to exert active
forces on the surrounding liquid and to set it in motion.  Conversely,
hydrodynamic friction forces acting along the flagellum feedback on
its motor dynamics and change speed and shape of the flagellar
beat. Previous research reported a decrease
of wave amplitude and frequency in swimming medium of increased viscosity~\cite{brokaw1966effects,brokaw1975effects}.  
A force-velocity relationship of the flagellar beat for a time-varying load has been
measured in a swimming alga by exploiting rotational self-motion of
the cell~\cite{geyer2013cell}.  
Additionally, external fluid flow can perturb the shape of the beat~\cite{wan2014rhythmicity}. 
These experiments indicate an intricate flagellar load-response, which
represents a cell-scale counterpart of the known force-velocity
relations of individual molecular motors~\cite{hunt1994force} or the
torque-speed relation of the bacterial rotary motor~\cite{chen2000torque}.

The ability of the beating flagellum to respond to external forces forms
the basis of the remarkable phenomenon of flagellar synchronization by
mechanical coupling~\cite{Taylor:1951,brumley2014flagellar}. On
epithelial surfaces, collections of cilia phase-lock to a common beat
frequency~\cite{sanderson1981ciliary} despite active
noise~\cite{polin2009chlamydomonas,ma2014active,goldstein2009noise}.
The result are metachronal waves of coordinated cilia beating, which
facilitate efficient fluid transport~\cite{elgeti2013emergence}.
Pairs of flagella can synchronize their beat, \textit{e.g.} in the
bi-flagellated green alga \textit{Chlamydomonas}, which constitutes a
model system for the study of flagellar synchronization
~\cite{ruffer1987comparison,goldstein2009noise,geyer2013cell}.  Here,
we present a novel and versatile approach that allows us to predict
the response of flagellar bending waves to external flows from
recordings of their unperturbed dynamics, using \textit{Chlamydomonas} 
as model system.

We first characterize how flagellar bending waves change phase speed
and amplitude in response to external forces, by exposing
\textit{Chlamydomonas} cells to controlled homogeneous flows.  We
complement these experiments by a theoretical description of the
flagellar beat as a generic limit-cycle oscillator, which may be
considered a realistic analogue of previous idealized models that
represented the beating flagellum by a sphere revolving around a
circular orbit \cite{Vilfan:2006,niedermayer2008synchronization,uchida2011generic,friedrich2012flagellar}.
% whose motion was likewise characterized by phase and amplitude. 
This theory, parameterized using measurements in the absence of
flow, predicts deviations of perfect limit-cycle dynamics in the
presence of external perturbations in quantitative agreement with
experiments.  Our theory further allows to computationally predict 
% the response of a beating flagellum to time-varying flows, \textit{e.g.} 
flagellar synchronization in external oscillatory flows.

\paragraph{Experimental setup.}
Single cells were held in a micropipette in a region of homogeneous flow inside a microfluidic
channel, see Fig.~1(a). High-speed recordings (1 kHz) of flagellar
beating were performed for different external flow velocities $u$,
see Supplemental Material (SM) for further details \cite{somrefs}. % on the experimental protocol.
For all cells analyzed, both flagella were beating in synchrony ${\ge}80\%$ of
the time, with a common frequency of $f_0 = 49.3 \pm 5.7\,\mathrm{Hz}$
when no external flow was applied ($\text{mean}{\pm}\text{s.e.}$,
$n{=}6$).  The shape of flagellar centerlines was tracked using custom
software and is fully characterized by its tangent angle profile
$\psi(s,t)$ as a function of arclength $s$ and time $t$
\cite{riedel2007molecular}, see Fig.~1(b). 
We limited our analysis to datasets with successful tracking of ${\ge}80 \%$ of the flagellar
length in ${\ge}80 \%$ of all frames for one flagellum,
corresponding to the unbiased selection of $n{=}6$ high-quality datasets.
To illuminate the load-response of the flagellar beat during different phases of its
beat, we first need to introduce the concept of flagellar
phase~\cite{goldstein2009noise,geyer2013cell,wan2014rhythmicity}. 

\paragraph{Limit-cycle reconstruction.}
We map the periodic dynamics of the flagellar tangent angle
$\psi(s,t)$ on a complex oscillator variable $Z{=}A\exp(i\varphi)$ with
oscillator phase $\varphi$ and normalized amplitude $A$, see
Fig.~1(c), using a limit-cycle reconstruction technique
~\cite{ma2014active,werner2014shape}. The logic is that in the
absence of intrinsic fluctuations and external perturbations,
flagellar oscillations can be characterized by 
a phase $\varphi$ that increases at a constant rate equal
to the angular frequency of the flagellar beat,
$\dot{\varphi}=2\pi\, f_0$,
and constant amplitude $A=A_0$. 
In short, we used principal component
analysis to express measured tangent angle profiles as superposition
of two fundamental shape modes,
$\psi(s,t)\approx \psi_0(s) +\beta_1(t)\psi_1(s) +
\beta_2(t)\psi_2(s)$, which was followed by a normalization step that
defines $Z(t)$ in terms of $Z_\mathrm{raw}(t)=\beta_1(t)+i\beta_2(t)$.
Specifically, we find that $Z_\mathrm{raw}(t)$ jitters around a limit
cycle $\ol{Z}_\mathrm{raw}(\varphi)$.  This limit-cycle can always be
parametrized by a phase $\varphi$ such that the mean phase speed is
independent of $\varphi$~\cite{Kralemann2008}.  We assign a unique
phase $\varphi$ to every flagellar shape by projecting its `shape
point' $\Z_\mathrm{raw}(t)$ radially onto this limit-cycle.  The
normalized amplitude $A=|Z_\mathrm{raw}|/|\ol{Z}_\mathrm{raw}|$ gauges
deviations from the limit-cycle in the radial direction.
Phase and amplitude allow a faithful reconstruction of waveform, 
see Fig.~S4 in SM.
We used the same set of shape modes for all data-sets:
this enables a direct comparison of waveform changes among different flagella.

\paragraph{External flow changes speed and amplitude of flagellar oscillations.}
In the presence of external flow with velocity $u$, 
we observe consistent changes of flagellar bending waves.
Fig.~1(d) shows the change in phase speed $\dot\varphi$ under external flow $u>0$:
at $\varphi=0$ (during the effective stroke), the beat speeds up, 
while it slows down at $\varphi=\pi$ (recovery stroke). 
The recovery stroke approximately lasts $0.80{\cdot}\pi-1.75{\cdot}\pi$
(corresponding to the part of the beat cycle, 
during which the center-of-mass of the flagellum moves up, relative to the long axis of the cell body).
Next, for each value of the phase $\varphi$, we performed a linear regression of
instantaneous phase speed versus flow velocity,
$\dot\varphi \approx 2\pi f_0\left(1 + \chi_\varphi\,u\right)$,
yielding two fit parameters $f_0$ and $\chi_\varphi$, see~Fig.~1(d).
We thus obtained a phase-dependent susceptibility
$\chi_\varphi(\varphi)$ for the phase speed, and similarly
$\chi_A(\varphi)$ for the amplitude.
\begin{figure}[h!]
  \centering
  \includegraphics[width=0.5\textwidth]{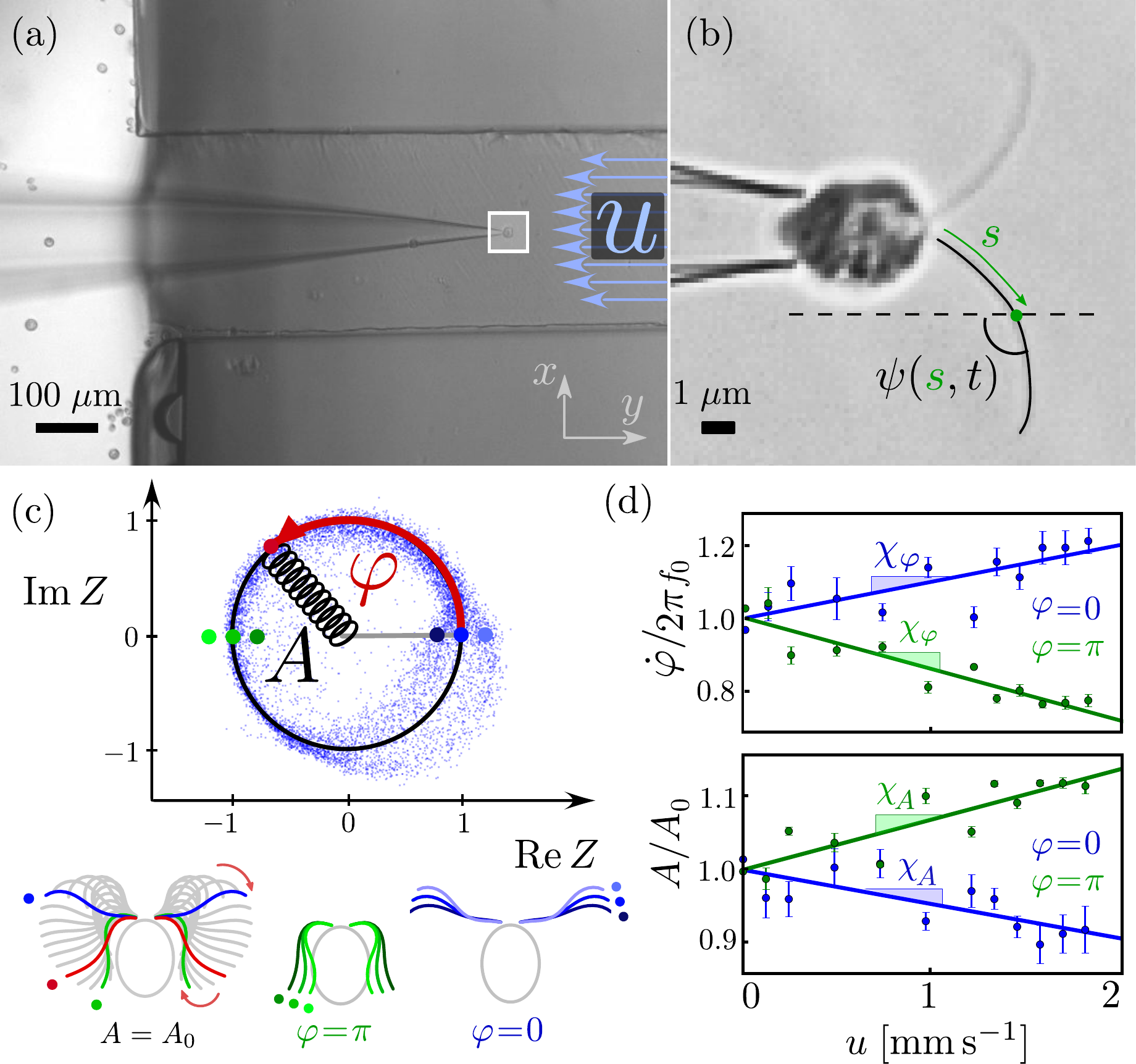}
  \caption{ 
    \textit{
    Load-response of the flagellar beat to controlled
    microfluidic flows.}  (a) \textit{Chlamydomonas} cell held in a
    micropipette exposed to flow in a microfluidic channel (velocity $u$
    at channel midline).  (b) Flagellar shapes and their tangent angle
    profiles $\psi(s,t)$.  (c) We map each flagellar shape on a shape
    point $Z=A\exp(i\varphi)$, see text.  This defines an
    instantaneous phase $\varphi$ and amplitude $A$ of the flagellar
    beat, as well as a limit-cycle of flagellar oscillations (black)
    that averages out fluctuations and measurement noise.  Example
    flagellar shapes are shown for reference shape points (colored
    dots). (d) Susceptibilities $\chi_\varphi$ and $\chi_A$
    characterize the change of instantaneous phase speed and amplitude
    as function of external flow $u$. 
    Error bars=s.e.m.\ for 5 sub-datasets. 
}
  \label{fig:data}
\end{figure}

\paragraph{Generic theory of flagellar oscillations.}
We present a generic theory to describe flagellar bending waves,
which is independent of specific assumptions on the detailed
mechanisms of motor control inside the flagellum, to predict the
measured susceptibilities $\chi_{\varphi}\left(\varphi\right)$ and
$\chi_A\left(\varphi\right)$. We build on the basic fact that regular
flagellar bending waves are stable in time, \textit{i.e.}  represent a
limit-cycle oscillator. 
We will stick to the simple, one-dimensional case of a scalar amplitude, inline with our analysis
  of experimental data. 
Thus, we can describe flagellar dynamics of
a \textit{Chlamydomonas} cell by a minimal number of degrees of
freedom. We delimit ourselves to the particularly simple case of
perfectly symmetric beating. Thus, for a freely swimming cell, the
phase space vector reads $\q=(\varphi,A,y)$.  Here, $\varphi$ and $A$
denote phase and amplitude of the flagella, while $y$ denotes the
position of the cell body center, where we chose the $y$-axis parallel
to the long axis of the cell body, see~Fig.~1(a,b).  By symmetry, the
cell swims along the $y$-direction.

We now derive equations of motion for the swimming cell from a balance of forces between
active flagellar driving forces $Q_j(\varphi)$ and friction forces
$P_j(\varphi)$, using a general framework of Lagrangian mechanics for
dissipative systems~\cite{polotzek2013three}.
The total rate of dissipation $\mathcal{R}_\mathrm{tot}$ 
of the system consisting of flagella and surrounding fluid 
can be written as
\begin{equation}
  \label{eq:dissipationrate}
  \mathcal{R}_\mathrm{tot} = \mathbf{P}\cdot\dot{\q} 
= P_\varphi\dot{\varphi} + P_A\dot{A} + P_y\dot{y},
\end{equation}
where the generalized friction forces $P_j$
are conjugate to the generalized velocities $\dot q_j$,
$j\in\{\varphi,A,y\}$. 
For example, $P_\varphi\dot{\varphi}$ would be the rate of dissipation
associated with advancing both flagella in the beat cycle at speed $\dot{\varphi}$, 
while keeping all other degrees of freedom fixed.
Friction forces will comprise both a
contribution from hydrodynamic friction of the surrounding viscous
fluid, as well as possibly internal friction inside the beating
flagella.

\begin{figure}[h!]
  \centering
  \includegraphics[width=0.5\textwidth]{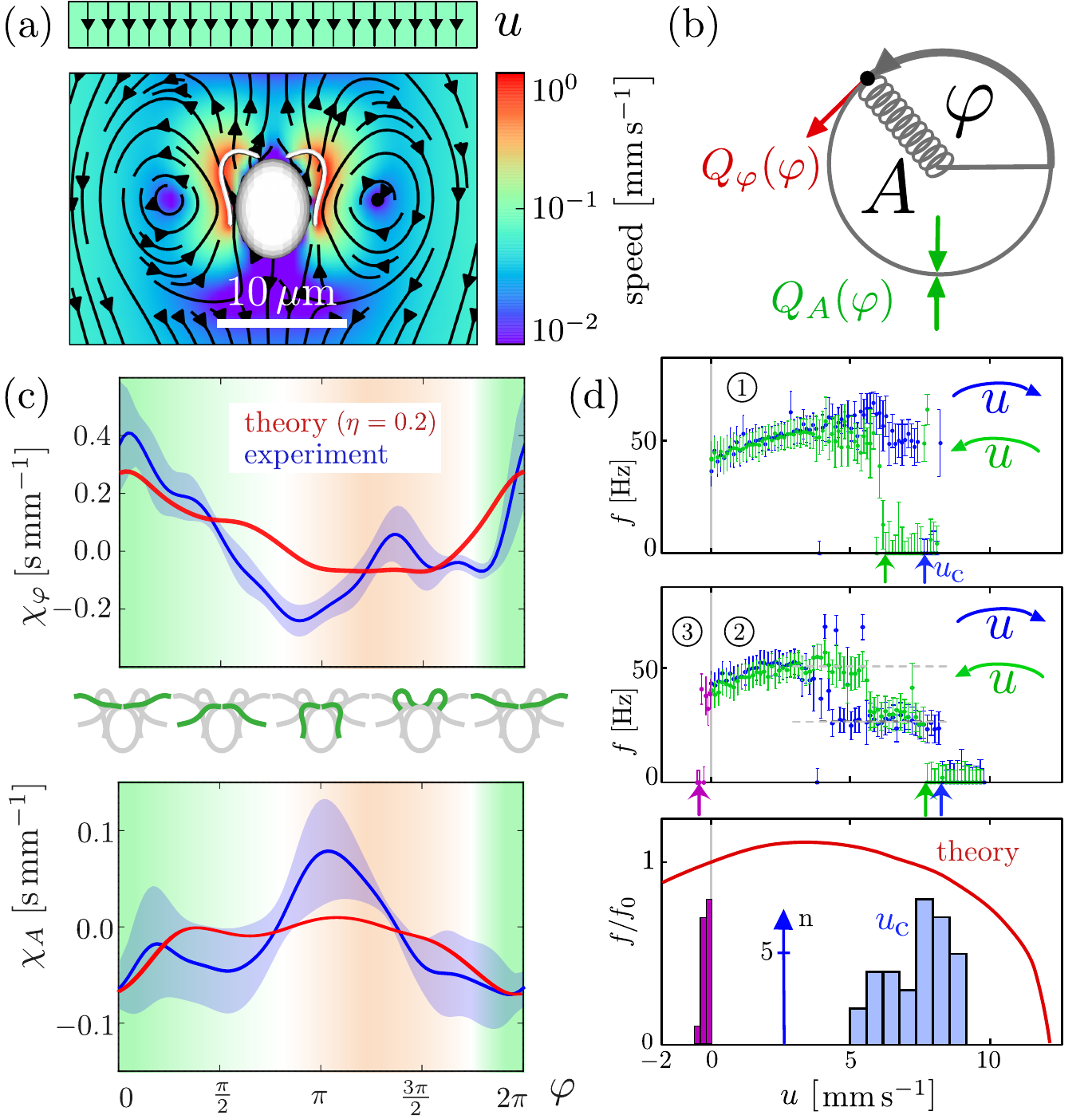}
  \caption{ \textit{Theory and experiment of flagellar load-response.}
    (a) We computed hydrodynamic friction forces acting on a beating
    flagellum numerically by solving the Stokes equation for viscous
    flow.  (b) We introduce a generic theoretical description of
    flagellar oscillations as a limit-cycle oscillator with phase
    $\varphi$ and amplitude $A$.  An active flagellar driving force
    $Q_{\varphi}$ coarse-grains active motor dynamics inside the
    flagellum.  A constraining force $Q_A$ for the amplitude ensures
    stability of the limit-cycle.  These phase-dependent active forces
    are uniquely calibrated from experimental data in the absence of
    flow.  (c) Phase-dependent susceptibilities $\chi_\varphi$ and
    $\chi_A$ for phase speed and amplitude, respectively.
    Comparison of experiment (blue, mean$\pm$s.e.) and theoretical
    prediction (red) yields $\efficiency=0.21\pm 0.06$ (mean$\pm$s.e.,
    $n{=}6$) for the energy efficiency of the flagellar beat.
	The duration of effective and recovery stroke are indicated by green and red shading, respectively.
    (d) External flow induces stalling of flagellar oscillations beyond a critical
    flow velocity $u_c$.  
    \textit{Top,middle:} 
    typical frequency responses for three flagella, numbered 1,2,3.    
    \textit{Bottom:} theory prediction of
    frequency response and stalling (red); inset: histograms of
    experimentally measured $u_c$ ($u{>}0$: blue, $n{=}33$ flagella, $u{<}0$: lilac, $n{=}16$ flagella).  }
\label{fig:model}
\end{figure}

Any flagellar dynamics will set the surrounding fluid in motion, and
will thus induce viscous dissipation.  In the overdamped regime of low
Reynolds number, applicable to cellular
swimmers~\cite{lauga2009hydrodynamics}, hydrodynamic friction forces
are linearly related to the velocities, \textit{i.e.}
$\P^{(h)} = \mathbf{\Gamma}^{(h)} \cdot \dot{\q}$.  The hydrodynamic
friction matrix $\mathbf{\Gamma}^{(h)}(\mathbf{q})$ can be calculated
by solving the Stokes equation with boundary conditions on the surface
of the swimmer specified by $\q$ and $\dot\q$; Fig.~\ref{fig:model}(a)
shows an example with $\q = (\pi, 1, 0)$ and
$\dot\q = (2\pi f_0, 0, u)$. 
In principle, these generalized friction forces can be
computed for any beat pattern parameterized by phase $\varphi$ and
amplitude $A$. Here, we employ the beat pattern shown in
Fig.~\ref{fig:data}(c) and a fast boundary element
method~\cite{liu2006fast,klindt2015flagellar}. As technical
side-note, in simulations involving large flow velocities (\textit{e.g.}
for flagellar stalling studied below) the range of the
flagellar amplitude was constrained to account for steric interactions
between flagella and cell body.

In addition to hydrodynamic dissipation in the surrounding fluid,
energy is also dissipated inside the beating flagellum itself.  This
intraflagellar friction combines effects of incomplete energy
conversion by molecular motors and possibly dissipation by
viscoelastic structural elements in the flagellar axoneme such as
nexin linkers.  In the absence of detailed knowledge on this
intraflagellar friction, we make the simple ansatz of a
rate of intraflagellar dissipation $\mathcal{R}^{(i)}$ 
proportional to the total dissipation rate 
\begin{equation}
\label{eq:energyefficiency}
\R^{(i)} = (1 - \efficiency) \R_{\mathrm{tot}}.
\end{equation}
Here, $\eta$ represents an efficiency of chemo-mechanical energy conversion.
The total friction matrix can then be written as
$\Gamma_{ij} = \Gamma_{ij}^{(h)}/\efficiency$, with
$i, j \in \{\varphi, A\}$ and the total friction forces read
$\P=\mathbf{\Gamma}\cdot\dot{\q}$.

The equation of motion of flagellar dynamics is specified by a balance of generalized
forces
\begin{equation}
  \label{eq:forcebalance}
  Q_\varphi = P_\varphi \text{ and } Q_A = P_A,
\end{equation}
where the active driving forces $Q_\varphi$ and $Q_A$ conjugate to
$\varphi$ and $A$ coarse-grain active motor dynamics inside the
flagellum, see also Fig.~\ref{fig:model}(b).  Below, we calibrate
these active driving forces using measured limit-cycle dynamics in the
absence of flow.  The force $P_y$ conjugate to $y$ corresponds to the
$y$-component of the total force acting on the cell. For a freely
swimming cell, $P_y{=}0$.  For a clamped cell, $y$ is constant and
$P_y$ represents the constraining force required to ensure this
constraint.  The case of an external flow with constant velocity $u$
is incorporated in this formalism by setting $\dot{y}=u$.

By convention, the unperturbed limit-cycle dynamics of the flagellar
beat for $u=0$ is characterized by $\dot\varphi = 2\pi\, f_0$ and
$A = A_0$.  This requirement uniquely determines $Q_\varphi(\varphi)$
and $Q_A(\varphi,A)$ for $A=A_0$,
as 
$Q_\varphi(\varphi)=\Gamma_{\varphi\varphi}(\varphi,A_0)2\pi f_0$ and
$Q_A(\varphi,A_0)=\Gamma_{A\varphi}(\varphi,A_0)2\pi f_0$.
Next, we extend the domain of definition of $Q_A$ to the case $A\neq A_0$ 
by making the simple ansatz of an effective amplitude spring acting on $A-A_0$
\begin{equation}
  \label{eq:amplitudeansatz}
Q_A(\varphi,A) = Q_A(\varphi, A=A_0) - k_A(\varphi)(A-A_0).
\end{equation}
This amplitude spring with spring constant $k_A$ ensures the stability
of flagellar limit-cycle oscillations with respect to amplitude
perturbations, which has been observed
experimentally~\cite{wan2014rhythmicity}.  In a simple
phenomenological description, amplitude perturbations decay with a
characteristic relaxation time $\tau_A$, $\tau_A\, \dot A = A_0-A$.
For simplicity, we assume that $\tau_A$ does not depend on phase
$\varphi$.  A value of $\tau_A=5.9\, \mathrm{ms}$ has been previously
determined by a measurement of the correlation time of amplitude
fluctuations~\cite{ma2014active}.  Equation
(\ref{eq:amplitudeansatz}) uniquely specifies a choice for the
phase-dependent amplitude stiffness $k_A(\varphi)$, 
see SM.
All quantities in our theoretical description of flagellar limit-cycle dynamics are now calibrated, 
except for the unknown flagellar efficiency parameter $\efficiency$,
where we used only beat patterns measured in the absence of external flow.  
Next, we compare theory and
experiment for the load-response of flagellar oscillations in the
presence of external flow, and determine $\eta$ by a simple fit.

\paragraph{The load-response of the beating flagellum.}  
We characterize the load-response of the flagellar beat to external
flow in terms of phase-dependent susceptibilities
$\chi_\varphi(\varphi)$ and $\chi_A(\varphi)$ for phase speed and
amplitude, respectively, as introduced already in Fig.~1(d). 
An average of $\chi_\varphi$ for $n{=}6$ cells is shown in
Fig.~\ref{fig:model}(c), revealing a reproducible response of
flagellar bending waves to external flow. 
During the effective stroke ($\varphi \approx 0$), we observe an increase in phase speed
($\chi_\varphi>0$), whereas the recovery stroke slows down
($\chi_\varphi<0$ for $\varphi \approx \pi$). 
These two effects partially cancel, resulting in a net change of the beat frequency by a
few percent only, see also Fig.~2(d).  
% Note that the cycle-averaged phase speed equals the angular beat frequency $2\pi f$ by definition.
Our analysis highlights the importance of a sub-cycle analysis of the
flagellar load-response. For the amplitude, we find $\chi_A > 0$ for
$\varphi\approx\pi$, \textit{i.e.} the flagella are closer to the
cell body during the recovery stroke. 
Our simple theory captures the principal features of the experimentally determined
phase-dependence, yet not all details can be reproduced.
A one-parameter fit provides an estimate for
the flagellar efficiency parameter $\eta$, yielding
$\efficiency = 0.21\pm 0.06$ (mean$\pm$s.e., $n{=}6$).
The flagellar susceptibilities used to estimate $\eta$ are robust with respect to uncertainty in $\tau_A$ 
or using a different waveform, see Figs.~S5 and S6 in SM.

\paragraph{Stalling of flagellar oscillations.}
For strong external flow, we find that the recovery stroke slows down
both in theory and experiment -- up to a point, where the flagellar
beat comes to a halt.  While increasing flow velocity $u$ in experiments,
we consistently observed first a slight increase of beat frequency
$f$, followed by a rapid decrease; two generic cases are shown in
Fig.~\ref{fig:model}(d).  Above a critical flow velocity $u_c$, the
flagellar beat arrests in the recovery stroke. %, corresponding to $\dot{\varphi}{=}0$.  
Upon decreasing $u$, flagella reproducibly resumed regular beating. 
Occasionally, a slight hysteresis was noticeable, see \textit{e.g.} Fig.~\ref{fig:model}(d)-\raisebox{.5pt}{\textcircled{\raisebox{-.9pt} {1}}}. 
In 10 out of 16 \textit{cis}-flagella (yet not in \textit{trans}-flagella), we observed a different, non-planar mode of flagellar beating 
with distinctly different frequency for intermediate values of $u$, see \textit{e.g.}~\mbox{Fig.~\ref{fig:model}(d)-\raisebox{.5pt}{\textcircled{\raisebox{-.9pt} {2}}}}
and Supplemental Movie S1.
Otherwise, we did not observe significant differences in the load-response of \textit{cis}- and \textit{trans}-flagella.
For negative flow velocities, the flagellar beat stalls already at $u\approx -0.5\,\mathrm{mm/s}$,
see \textit{e.g.} Fig.~\ref{fig:model}(d)-\raisebox{.5pt}{\textcircled{\raisebox{-.9pt} {3}}} and Supplemental Movie S2.

Our simple theory predicts a similar dependence of beat frequency $f$ on flow velocity for $u{>}0$,
with stalling at a critical flow velocity $u_c=14\,\mathrm{mm}\,\mathrm{s}^{-1}$, see Fig.~2(d). 
This flagellar stalling displays characteristic signatures of a saddle-node bifurcation. 
Intriguingly, our simple theory predicts flagellar stalling for $u{>}0$ much better than for $u{<}0$, 
when stalling occurs during the effective stroke, characterized by low flagellar curvature.
Flagellar curvature has been previously proposed as key determinant of flagellar motor control \cite{brokaw1971bend,sartori2016dynamic}.
We anticipate that flagellar load-responses and stalling can provide a test case for refined theories of flagellar beating
\cite{brokaw1971bend,lindemann1994geometric,camalet2000generic,riedel2007molecular,sartori2016dynamic}.

\begin{figure}[h!]
  \centering
  \includegraphics[width=0.45\textwidth]{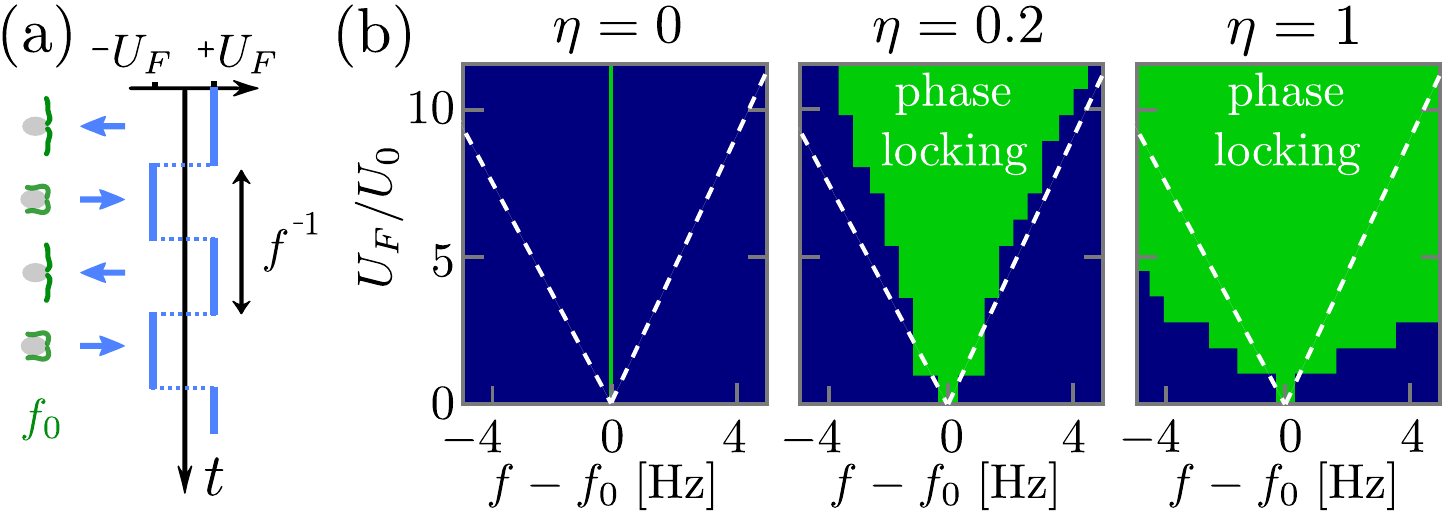}
  \caption{
  \textit{
Phase-locking of \textit{Chlamydomonas} flagellar beat to external oscillatory flow.
}
(a) 
Sketch of original experiment \cite{quaranta2015hydrodynamics}.
(b) 
Computed Arnold tongues as function of  flow frequency $f$ and amplitude $U_F$ 
for different values of the flagellar energy efficiency $\efficiency$.
Dashed lines indicate the phase-locking region measured in \cite{quaranta2015hydrodynamics}.
Parameters: $U_0 = 0.11\,\mathrm{mm\,s^{-1}}$, $f_0 = 50\,\mathrm{Hz}$ intrinsic frequency of flagellar beat.}
\label{fig:implications}
\end{figure}

\paragraph{Flagellar phase-locking to external driving.}
Recent experiments by Quaranta \textit{et al.} demonstrated
phase-locking of \textit{Chlamydomonas} flagellar beating to external
oscillatory fluid flow \cite{quaranta2015hydrodynamics}. In
accordance, our theory predicts similar phase-locking, see
Fig.~\ref{fig:implications}.  The width of the resultant Arnold tongue
agrees with measurements from \cite{quaranta2015hydrodynamics} within
$20\%$ (using $\eta{=}0.2$).  
Naturally, this flagellar synchronization 
relies on the fact that beating flagella
respond to changes in hydrodynamic load. 
The width of the Arnold tongue would approach zero for 
$\chi_\varphi{=}\chi_A{=}0$,
\textit{i.e.} for a flagellum with vanishing energy efficiency $\efficiency{=}0$.  
On the other hand, assuming maximal efficiency $\eta{=}1$ would result in an
Arnold tongue that is too wide to account for the experimental
observations.
Likewise, synchronization of \textit{cis}- and \textit{trans}-flagellum
depend on flagellar load-responses \cite{goldstein2009noise,geyer2013cell}; 
for this, phase-dependent driving forces and non-isochrony of flagellar oscillations (see Figs.~S3 and S8 in SM) are relevant \cite{niedermayer2008synchronization,uchida2011generic}.

\paragraph{Discussion.}
% We are only beginning to understand the self-organized dynamics in
% collections of molecular motors and how these respond to external
% forces~\cite{riedel2007molecular,pelle2009mechanical,lindemann1988calcium,julicher1997modeling}.
We report direct measurements of dynamic load-responses in an
evolutionary conserved motor system, the beating flagellum. We
characterized the flagellar beat as a limit-cycle oscillator that
displays characteristic phase and amplitude susceptibilities in
response to an external force. Motivated by this description, we
formulated a theory of flagellar dynamics in terms of a
phase-amplitude oscillator, which is calibrated by experimental data
in the absence of external flow.
This minimal theory, based on the assumption of small perturbations of flagellar limit cycle oscillations, is able to predict key features of the flagellar load-response, such as acceleration of the effective stroke and deceleration of the recovery stroke by external flow, as well as reversible stalling of flagellar oscillations beyond a critical flow velocity, in quantitative agreement with experiment. Details of the phase-dependent load-response, hysteresis of flagellar stalling, switching to a second, non-planar mode of beating, as well as stalling at surprisingly small negative flow velocities, are not accounted for by our simple theory, yet may be informative for refined theories of flagellar beating and the underlying dynamics of molecular motors 
\cite{brokaw1971bend,lindemann1994geometric,camalet2000generic,riedel2007molecular,sartori2016dynamic}. 

% \paragraph{Energetics of the flagellar beat.} 
By comparison of theory and experiment, we obtained an estimate 
$\eta=0.21\pm 0.06$ 
for the energy efficiency of the flagella beat, 
defined as the ratio of mechanical work
performed by the beating flagellum on the surrounding fluid divided by the
chemical energy input required to sustain its beat. 
Previous estimates for $\eta$ reported values in the range
$\efficiency = 0.1 - 0.4$~\cite{brokaw1967adenosine,katsu2009substantial,Cardullo:1991,chen2015atp},
see also SM.
In our theory, we assume an active flagellar driving force that is
independent of external load.  This assumption is consistent with
recent experiments, which show that flagellar ATP consumption is rather insensitive to beat frequency and mechanical
load~\cite{chen2015atp}.  Accordingly, any increase in load is
compensated by a reduction in speed, not an increase in fuel
consumption.  

% \paragraph{Implications for synchronization.} 
The load-response of beating flagella reported here 
is an essential prerequisite % has direct implications 
for flagellar synchronization by mechanical coupling, \cite{Friedrich2016},
without which hydrodynamic synchronization would be impossible.

% The flagellar load-response represent an instance of fluid-structure interactions for an active, elastic structure, the beating flagellum. 
% Obviously, a hypothetical flagellum with zero 
% susceptibility to external mechanical forces could not entrain to any periodic mechanical driving. 
% Additionally, Lenz \textit{et al.} emphasized 
% the effect of amplitude compliance for flagellar synchronization by a coupling of amplitude and phase speed~\cite{niedermayer2008synchronization}.
% Here, we directly quantified this active amplitude compliance for a beating flagellum.
% As an alternative synchronization mechanism, Golestanian \textit{et al.}
% discussed the role of phase-dependent active driving forces~\cite
% {uchida2011generic,golestanian2011hydrodynamic}.
% Here, we reconstructed an effective flagellar driving force and found indeed 
% that it varies during the
% flagellar beat cycle.
% We envision that, in addition to synchronization, also other dynamic phenomena such as
% the interaction between a beating flagellum and a boundary
% wall~\cite{denissenko2012human}, or active reorientation responses of flagellated
% microswimmers in external shear flows~\cite{miki2013rheotaxis}
% depend sensitively on the load-response and waveform compliance of the beating flagellum.

% \paragraph{Acknowledgments.}
\begin{acknowledgments}
G.S.K. and B.M.F. acknowledge financial and institutional support from
the German Science Foundation ``Microswimmers" priority program SPP
1726 (Grant No. FR 3429/1-1).  We thank E. Terriac for help with
design and fabrication of microfluidic chips and useful discussions.
G.S.K and C.R. contributed equally to this work.
\end{acknowledgments}
\nocite{gorman1965cytochrome}
\nocite{leptos2013antiphase}
\nocite{matpiv}

\bibliography{refs}{}
\bibliographystyle{steffensapsrev}

\end{document}